\documentclass[twocolumn,prl,aps,showpacs]{revtex4b4}
\usepackage{epsfig}

\begin{document}
\newcommand{\be}{\begin{equation}}
\newcommand{\bea}{\begin{eqnarray}}
\newcommand{\ee}{\end{equation}}
\newcommand{\eea}{\end{eqnarray}}
\def \blank{\mbox{}}
\def\d{\mbox{$\partial$}}
\def\NB{\mbox{$N_{bin}$}}
\def\de{\mbox{$d_E$}}
\def\dw{\mbox{$d_W$}}
\def\dl{\mbox{$d_L$}}
\def\da{\mbox{$d_A$}}
\def\hb{\mbox{$\hbar$}}
\def\Re{\mbox{\,Re\,}}
\def\Im{\mbox{\,Im\,}}
\def\dw{\mbox{$d_W$}}
\def\nb{\mbox{$N_B$}}
\def\A{\mbox{\bf A}}
\def\B{\mbox{\bf B}}
\def\C{\mbox{\bf C}}
\def\M{\mbox{\bf M}}
\def\P{\mbox{\bf P}}
\def\x{\mbox{{\bf x}}}
\def\ol{\mbox{$\bar{\lambda}$}}
\def\Q{\mbox{{\bf Q}}}
\def\R{\mbox{{\bf R}}}
\def\S{\mbox{{\bf S}}}
\def\L{\mbox{$L$}}
\def\Rd{\mbox{$R^d$}}
\def\DF{\mbox{{\bf DF}}}
\def\DG{\mbox{{\bf DG}}}
\def\DH{\mbox{{\bf DH}}}
\def\DFL{\mbox{$\DF^L$}}
\def\DGL{\mbox{$\DG^L$}}
\def\OSL{\mbox{{\bf OSL}}}
\def\T{\mbox{$T$}}
\def\Ec{\mbox{{$\cal E$}}}
\def\Pc{\mbox{{$\cal P$}}}
\def\Ac{\mbox{{$\cal A$}}}
\def\l{\mbox{$\lambda$}}
\def\g{\mbox{$\gamma$}}
\def\oma{\mbox{$\omega_A$}}
\def\om{\mbox{$\omega$}}
\def\eps{\mbox{$\epsilon$}}
\def\Y{\mbox{$\bf{Y}$}}
\def\y{\mbox{$\bf{y}$}}
\def\v{\mbox{$\bf{y}$}}
\def\ynn{\mbox{$\bf{y}^{NN}$}}
\def\w{\mbox{$\bf{w}$}}
\def\z{\mbox{$\bf{z}$}}
\def\e{\mbox{$\bf{e}$}}
\def\del{ \mbox{\boldmath{$\Delta$}} }
\def\delt{ \mbox{\boldmath{$\delta$}} }
\def\domega{ \mbox{$\delta\omega$}}
\def\xint{\mbox{$\displaystyle\int d^3x \,$}}
\def\f{\mbox{$\bf{f}$}}
\def\p{\mbox{$\bf{\rho}$}}
\def\F{\mbox{$\bf{F}$}}
\def\E{\mbox{$\bf{E}$}}
\def\D{\mbox{$\bf{D}$}}
\def\H{\mbox{$\bf{H}$}}
\def\G{\mbox{$\bf{G}$}}
\def\U{\mbox{$\bf{U}$}}
\def\J{\mbox{$\bf{J}$}}

\title{Distribution of Mutual Information}

\author{Henry D. I. Abarbanel}
  \altaffiliation{Institute for Nonlinear Science}
  \email{hdia@jacobi.ucsd.edu}
  \affiliation{Department of Physics and Marine Physical Laboratory,
Scripps Institution of Oceanography, University of California, San Diego,
La Jolla, CA  92093-0402}

\author{Naoki Masuda} 
\email{masuda@physics.ucsd.edu}
\altaffiliation[Also ]{Department of Mathematical 
Engineering and Information Physics, Graduate School of Engineering, 
University of Tokyo, Tokyo, 113-8656, Japan}
\affiliation{Department of Physics and Institute for Nonlinear Science,
University of California, San Diego,La Jolla, CA  92093-0402}

\author{M. I. Rabinovich}
\email{rabin@landau.ucsd.edu}
\affiliation{Institute for Nonlinear Science,
University of California, San Diego, La Jolla, CA  92093-0402}

\author{Evren Tumer}
\email{evren@nye.ucsd.edu}
\affiliation{Department of Physics and Institute for Nonlinear Science,
University of California, San Diego, La Jolla, CA  92093-0402}

\date{\today}

\begin{abstract}

In the analysis of time series from nonlinear sources, mutual information 
(MI) is used as a nonlinear statistical criterion for the selection of an 
appropriate time delay in time delay reconstruction of the state space. MI 
is a statistic over the sets of sequences associated with the dynamical 
source, and we examine here the distribution of MI, thus going beyond the 
familiar analysis of its average alone. 
We give for the first time the distribution of MI for a standard, classical 
communications channel with Gaussian, additive white noise. For time 
series analysis of a dynamical system, we show how to determine the 
distribution of MI and discuss the implications for the use of average 
mutual information (AMI) in selecting time delays in phase space reconstruction.

\end{abstract}
\bigskip
\bigskip

\maketitle

Information theory~\cite{shannon,fano,gallagher} characterizes general
dynamical systems through a nonlinear connection between sequences of 
symbols associated with action of the system. The connection could be 
between inputs and outputs of a 
channel or could be associated with predictions of future measurements from past 
observations~\cite{schreib}. Shannon's identification of MI as the essential 
statistic in such systems gives a framework for the discussion of 
applications as diverse as fiber optic communications systems 
with time scales shorter than nanoseconds and nervous systems with 
time scales of a few tens of milliseconds to seconds. 

In the analysis of time series coming from nonlinear 
sources~\cite{abarbanel,kansch} one reconstructs a proxy state space 
using observations of a single variable $V(t)$. A useful 
reconstructed state space employs the observed variable $V(t)$ and its 
time delays to form a data vector
\be
\y(t)=[V(t),V(t-T\tau_s),  \ldots, V(t - (d-1)T\tau_s)],
\ee
where $\tau_s$ is the sampling time and $T$ is an integer. 
We must choose $T$ so that the components of 
$\y(t)$ are `independent enough' of each other to be 
good coordinates for the space. For this purpose
\bea
I(T) &=& \sum_{\{V(t),V(t+T\tau_s)\}} P(V(t),V(t+T\tau_s)) \nonumber \\
& &\log_2 \biggl\{\frac{P(V(t),V(t+T\tau_s))}{P(V(t))\,P(V(t+T\tau_s))} \biggr\},
\label{ami}
\eea
the AMI, 
has been used. Following the suggestion of Fraser and 
Swinney~\cite{andy}, $T$ is chosen where $I(T)$ has its first minimum.

The AMI $I(T)$ answers the question: how much, in bits, do we 
know about the measurement at $t + T\tau_s$ from the measurement at $t$, 
averaged over the whole time series or attractor. However, it does 
not tell us how MI is distributed over the time series. We have only the 
hope that the distribution of MI is sharp enough to provide a choice for 
$T$ which is useful over the whole time series. 

If the dynamics of the process producing $V(t)$ has two time 
scales, for example, one might expect the distribution of MI to reflect 
that by exhibiting two distinct peaks associated with each process 
efficiently connecting to itself through the 
dynamics of the system. In such a case the AMI may tell us little about how to 
select a value of $T$ to use in making $\y(t)$. Since the MI whose mean is
evaluated in Eq.(\ref{ami}) tells us how well we can predict $V(t)$ knowing
$V(t-T)$, the presence of two time scales leading to a multivalued 
distribution of MI is quite natural. 

In such circumstances, it may not be a good idea to use the $T$ 
chosen by the AMI, but to select different $T$ in different parts of 
the attractor, or to use another prescription altogether for choosing $T$. As many
physically or biologically interesting dynamical systems have two or more relevant
time scales, the distribution of MI presents an interesting issue.

Time delay phase space reconstruction is widely used as an initial step 
in the analysis of time series from nonlinear sources, so it is useful 
to establish how MI, considered as a `statistic' distributed over the 
measurements, is distributed. This will allow one to proceed well beyond the 
use of the average alone. 

Our considerations here may provide a framework for efficient use of a 
broad set of communications channels. Presented with a channel, which 
need not be stationary, one may wish to know those sequences which maximize 
MI to allow the best use of that channel. Those sequences need not 
always correspond to the AMI associated with the channel. In other words
the distribution of MI can depend both on the channel and on the symbol sequences
conveyed by the channel.

MI provides a nonlinear relation between one sequence of symbols and 
second sequence of symbols. These could be the input and output, 
respectively, of a communications system or of a network of neurons 
performing a functional task. We consider a sequence 
$S = \{...,s(-1),s(0),s(1),...\}$ observed at discrete times and ask 
about its nonlinear connection to a second sequence 
$R = \{...,r(-1),r(0),r(1),...\}$. The symbols 
$\{r(l)\}$ and $\{s(k)\}$ may take discrete or continuous values.

We are interested determining the properties of $S$ from measurements 
of the sequence $R$. The ability to do this is characterized by the MI
\be
\label{midef}
I(s(k),r(l))=\log\biggl\{\frac{P_{SR}(s(k),r(l))}{P_S(s(k))P_R(r(l))}\biggr\}.
\ee
$P_{SR}(s,r)$ is the joint distribution function of symbols $s$ taken 
from the input sequence $S$ and symbols $r$ taken from the output sequence $R$.
$P_{SR}(s,r)$ is the essential ingredient in determining MI. It 
depends both on the sequences $S$ and $R$ and, importantly, on the 
channel or dynamical process connecting the $s(k)$ to the $r(l)$. 
$P_S(s) = \sum_{r} P_{SR}(s,r)$ and 
$P_R(r) = \sum_{s} P_{SR}(s,r)$ are 
the distributions of $S$ symbols and $R$ symbols, respectively. 
When the sequences are independent, 
$P_{SR}(s,r) = P_S(s)P_R(r),\, \mbox{and} \, I(s,r)=0.$
The MI $I(s,r)$ is a variable over $(s,r)$, and it has its own 
distribution function $P_I(x)$. $P_I(x)$ tells us the 
frequency with which the value $x = I(s,r)$ occurs. 

The distribution of MI is defined by 
\be
P_I(x) = \int ds \, dr \, P_{SR}(s,r) \delta(x-I(s,r)),
\ee
with $I(s,r) = \log\biggl\{\frac{P_{SR}(s,r)}{P_S(s)P_R(r)}\biggr\}$.
$\langle x \rangle = \int dx \, x P_I(x) = \int ds \, dr \, P_{SR}(s,r) I(s,r)$,
as it should be. While $\langle x \rangle > 0$, negative values of $I(r,s)$
are possible~\cite{fano}. Negative values of $x = I(r,s)$ reflect the 
circumstance that
$(r,s)$ pairs may occur less frequently than the individual symbols 
themselves. If
a process produces out-of-phase appearances of $r$ and $s$, for example,
negative values of MI will occur. 

To evaluate $P_I(x)$ using the observed $P_{SR}(s,r)$ we could solve the 
delta function condition $x = I(s,r)$ for, say, $s_*(x,r)$ and then 
perform the integral over $r$. An alternative method is to Fourier 
transform $P_I(x)$ giving $Q_I(f) = \int dx \, e^{-2\pi i f x} P_I(x) = 
\int ds \, dr \, e^{-2 \pi i f I(s,r)} P_{SR}(s,r)$.
$P_I(x)$ is recovered by inverse Fourier transform
$P_I(x) = \int df e^{i2 \pi f x} Q_I(f)$.
In effect we are using the Fourier variable $f$ as a Lagrange 
multiplier to implement the required delta function on $x = I(s,r)$. 
We avoid solving for $s_*(r,x)$ and require only an integral over 
the same ingredients used in evaluating the AMI. 

A classical example is provided by the Gaussian distribution in $(r,s)$ 
\be
P_{SR}(s,r) = \frac{\sqrt{ab- \sigma^2}}{\pi} e^{-(ar^2+bs^2+2\sigma sr)}.
\ee
The quantity $\xi = \frac{\sigma^2}{ab} < 1$ for this to be 
normalizable. $P_R(r) = \sqrt{\frac{ab-\sigma^2}{\pi b}}e^{-r^2(1-\xi)}, \, P_S(s) =
\sqrt{\frac{ab-\sigma^2}{\pi a}}e^{-s^2(1-\xi)}$, and $I(s,r) = X_0 -(\frac{\sigma^2s^2}{a} + \frac{\sigma^2r^2}{b} + 2 \sigma rs)$, where $X_0 = -\frac{1}{2}\log(1-\xi)$. When $\sigma = 0, I(r,s) = 0.$

From this we find $Q_I(f)$ 
\be
Q_I(f) = e^{-i2 \pi f X_0} \frac{1}{\sqrt{1+4 \pi^2 f^2 \xi^2}}.
\ee
As expected, when $\xi = 0$, $Q_I(f) = 1$ and $P_I(x) = \delta(x)$. 
$P_I(x)$ can also be evaluated for 
$\xi \ne 0$
\be
P_I(x) = \frac{1}{\pi \xi} K_0\biggl(\frac{X_0 - x}{\xi}\biggr),
\ee
where $K_0(z)$ is a zeroth order modified Bessel function~\cite{absteg}. 
$K_0(z)$ is symmetric in $z$ and has a logarithmic 
singularity at $z = 0$.
The AMI is $X_0$, $\langle (x-X_0)^2\rangle = \xi^2$, and 
$\langle(x-X_0)^{2m}\rangle = \xi^{2m}((2m-1)!!)^2$.

A standard example of a Gaussian channel is given by the 
case of a Gaussian input signal with 
$P_S(s)=\frac{1}{\sqrt{2 \pi S}}e^{\frac{-s^2}{2S}}$, and with 
conditional probability of response
$P_{SR}(r|s) = \frac{1}{\sqrt{2 \pi N}}e^{\frac{-(r-s)^2}{2N}}$.
This is represents a statistical signal transported through a passive channel 
with additive, Gaussian, white noise~\cite{fano}. The signal to 
noise ratio in this channel is $\frac{S}{N}$.

The distribution of mutual  information is that just given with 
$a = \frac{1}{2N};\,b=\frac{1}{2N}(1+\frac{N}{S});\,
\sigma = \frac{-1}{2N}$, leading to the AMI 
$X_0 = \frac{1}{2} \log(1+\frac{S}{N})$, which is familiar \cite{fano}, 
and moments, which are a new result, given as above with 
$\xi = \frac{\frac{S}{N}}{1+\frac{S}{N}}.$ This means that for a large 
$S/N$, the AMI grows logarithmically in $S/N$, but, perhaps 
surprisingly, the moments about 
this average are order unity or larger. That is the distribution is 
not sharp, but becomes broad. For $\frac{S}{N}$ small, 
$X_0 \approx \frac{S}{2N}$ and the moments are powers of $\frac{S}{N}$. 
In fact, for small $\frac{S}{N}$ or noisy channels, the RMS value 
of MI is twice the mean value.

Other $P_{RS}(r,s)$ must be dealt with numerically, and the sequence 
values must be discrete. From $P_{SR}(s(k),r(l))$ a MI matrix is generated:
\be
\label{eq:Imat}
I(s(k),r(l)) = \log_2\left( \frac{P_{SR}(s(k),r(l))}{P_S(s(k)) \, P_R(r(l))}\right),
\ee
and $Q(f)$ is approximated by 
\bea
\label{eq:aprox_pk}
Q(f) &\approx&  \sum_{s(k),r(l)} P_{SR}(s(k),r(l)) 
e^{-i 2 \pi f I(s(k),r(l))}\Delta{r} \Delta{s} \nonumber \\
&=& \sum_{k} P_I(x_k) e^{-2 \pi i f x_k} \Delta x_k,
\eea
where $\Delta{s}$ and $\Delta{r}$ are the size of the symbol bins.  
$P_I(x)$ is evaluated bins of size $\Delta I$.
We denote $p_m = P_I(x_m) = P_I( m \Delta I + I_{min})$,
where m is an integer that ranges from $0$ to $\NB$ and 
$I_{min}$ is the minimum value of MI.  $p_m$, [\ref{eq:aprox_pk}] 
is determined for $\NB$ values of $f_n = \frac{n}{\NB \Delta I}$. $n$ is 
an integer in the range $[-\frac{1}{2}\NB,\frac{1}{2}\NB]$. The 
approximation to $Q(f_n)$ is then
$Q(f_n) = \Delta I \ e^{-2 \pi i I_{min} f_n} \ \sum_{k = 0}^{\NB} p_k \  
e^{-2 \pi i k \Delta I}$,
so the inverse Fourier transform of 
$\frac{Q(f_n)}{\Delta I} \exp\left(\frac{ 2 \pi i I_{min} n }{\NB\Delta I} \right)$ is $p_m$.

To apply this to the MI used in selecting the time delay in phase 
space reconstruction we identify $s(t) = V(t_0 + k\tau_s)=s(k)$ 
and $r(t) = V(t_0+(l+T)\tau_s)=r(l)$ where $t_0$ is some initial time 
in the data. We have investigated $P_I(x)$ for various $T$ 
for two simple dynamical systems: (1) the Lorenz system, and (2) 
a nonlinear hysteretic circuit using data provided by L. Pecora and 
T. Carroll (Personal Communication). Both data 
sets are discussed in~\cite{abarbanel}.

For the Lorenz system data set we used $2$x$10^6$ values of the 
variable $x(t)$. The first minimum of AMI for these data is 
at $T = 10$ in units defined by the numerical integration step. We 
evaluated $P_I(x)$ for various $T$. In Figure 1 
we show this distribution for $T=1$, for $9 \le T \le 11$, 
and for $T=16$. We can see that for $T=1$ $P_I(x)$ has a single well defined peak, 
indicating little variation of MI across the attractor. The AMI for 
this distribution is 
larger than that for $T \approx 10$. This is the usual~\cite{andy} indication that $T=1$ is not a preferred choice. For 
$9 \le T\le 11$ $P_I(x)$ is nearly the same with two well formed, but nearby, peaks. These we 
attribute to fast and slow motions on the attractor associated with motion 
about the unstable fixed points and motion through the ``neck" near 
the original in phase space. As the peaks are rather close, the use of a single $T$ to provide coordinates for the data vector $\y(t)$ over the whole attractor would appear to be a good construct. If one wished to resolve the two time
scales exposed by the peaks in $P_I(x)$, the use of two different
time delay coordinate systems in different parts
of the attractor would be appropriate.

 \begin{figure}[ht!]
\centerline{\psfig{figure=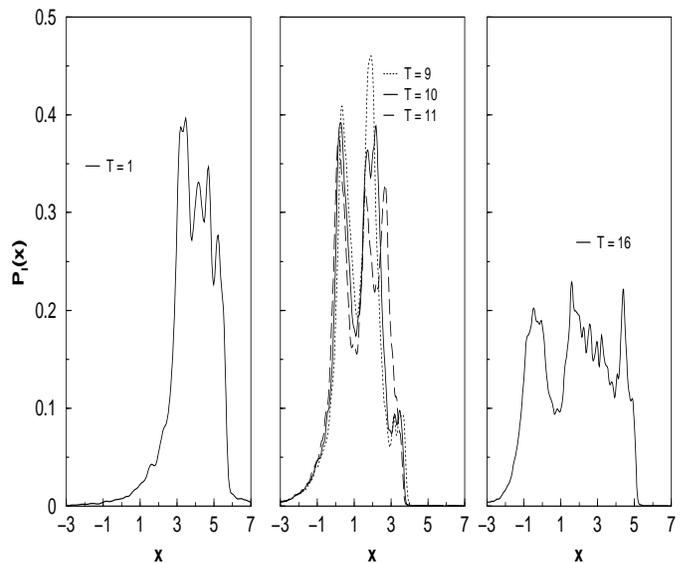,height=3.5in, width=3in,angle=270}
}
\caption{\small{Distribution of MI for the Lorenz system for various $T$ used in the
data vector $\y(t)$. The left panel has $T = 1$, the middle panel $9 \le T \le 11$, and the right
panel has $T = 16$. The AMI for $T = 1$ is larger than for 
$T \approx 10$. For $T =1$ and for $9 \le T \le 11$ we have rather sharp distributions 
$P_I(x)$, so the use of a single value of $T$ across the attractor is useful. 
For $T \approx 10$ the bimodal nature of $P_I(x)$ suggests that one could 
identify two processes contributing to $P_I(x)$ and seek a value for $T$ for 
each of them. For $T = 16$ the distribution is so broad that no particular
meaning can be associated with the AMI. This $P_I(x)$ indicates many different processes contribute over the attractor, and thus, this is not
a good value of $T$  to use in forming the data vector.}
}
\label{fig1}
\end{figure}

For larger $T$, here $T = 16$, we observe substantial broadening 
in $P_I(x)$. Here the mutual information between $V(t)$ and $V(t+T\tau_s)$ 
is more or less uniformly distributed over the attractor suggesting all points 
are rather decorrelated, in an information theoretic sense, from all others. 
This means that all dynamical processes on the attractor convey little information 
from time $t$ to time $t + T$. Equivalently, coordinates of the 
time delay data vector $\y(t)$ formed with $T = 16$ are so 
independent of each other as to constitute a bad choice for following 
the dynamics of the system.

For the nonlinear circuit data we have 64,000 data points taken at
$\tau_s = 0.1$\,ms. The minimum of AMI is $T = 6$. For these 
data $P_I(x)$ is shown in Figure 2 for $T=1$, for $5 \le T \le 7$, 
and for $T=11$. We have a relatively sharp peak for $T = 1$, but high values of AMI. 
The peak narrows for $T \approx 6$, and then broadens again 
as $T$ grows. We show $T = 11$ where the broadening of $P_I(x)$ seen already in the Lorenz system occurs again. This indicates that this value of $T$ is not useful in reconstructing the entire attractor.

\begin{figure}[ht!]
\centerline{\psfig{figure=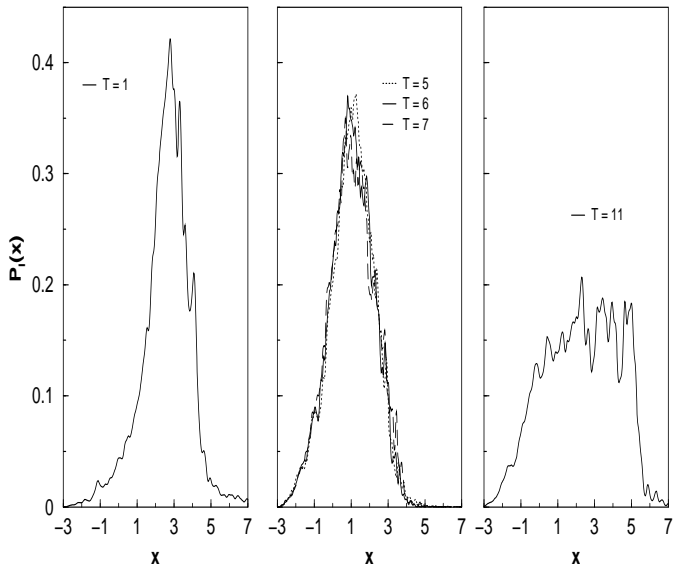,height=3.5in, width=3in,angle=270}
}
\caption{\small{Distribution of MI for the hysteretic circuit for various $T$ used in the
data vector $\y(t)$. The left panel has $T = 1$, the middle panel $5 \le T \le 7$, and the right
panel has $T = 11$. The AMI for $T = 1$ is larger than for $T \approx 6$. Each
of these is a rather sharp distribution $P_I(x)$, so the use of a single value of 
$T$ across the attractor is useful. For $T = 11$ $P_I(x)$ has broadened indicating 
this value of $T$ yields components of the data vector $\y(t)$ 
which may be too independent for use.}
 }
\label{fig2}
\end{figure}

$P_I(x)$ for various models and for a variety of experimental data 
will be reported in our larger paper~\cite{tumer1}. In this short 
note we have introduced the distribution of MI and exhibited several 
examples, analytic and numerical, to illustrate its properties. 
The use of $P_I(x)$ as illustrated here for understanding the distribution of MI in the connection between elements of the data vector $\y(t)$ gives us a clearer understanding of the choice made some years ago by Swinney and Fraser~\cite{andy} of using the first minimum of AMI to select $T$. It is not only low AMI among components of $\y(t)$ that is important, but also one must have a distribution of MI which is sharp so that a single choice for $T$ over the whole attractor can accurately capture the underlying dynamical processes.

We expect $P_I(x)$ to have direct utility in the study of networks with 
active dynamical elements~\cite{manuel,tumer1} where 
the function of the network is to convey information from a set of 
source symbols $s(k)$ to a set of response symbols $r(l)$ with 
high fidelity. This fidelity is characterized by high values of MI between 
input and output, and maxima of $P_I(x)$, rather than the AMI, 
will indicate which processes $s(k) \to r(l)$ are best 
communicated. These ideas will be fully explored in our 
larger paper~\cite{tumer1}.

\begin{acknowledgments}
This work was supported in part by the U.S. Department of Energy, 
Office Science, Division of Engineering and Geosciences, under 
grant DE-FG03-90ER14138, and in part by National Science Foundation grant 
NCR-9612250. Support was also received from the U. S. Army Research Office 
under contract No. DAAG55-98-1-0269; MURI Program in Chaotic Communications.
ET acknowledges support from NSF Traineeship DGE 9987614.
\end{acknowledgments}

\bibliography{infoprl2}
\bibliographystyle{revtex}

\end{document}